\documentclass[conference,letterpaper]{IEEEtran}

\usepackage{cite}
\usepackage{latexsym}
\usepackage{amsmath, amssymb}
\usepackage{mathrsfs}
\usepackage[dvips,dvipdf]{graphicx}
\usepackage{subfigure}
\usepackage{algorithm}
\usepackage{algorithmic}

\newcommand{\A}{\mathbf{A}}
\newcommand{\B}{\mathbf{B}}
\newcommand{\C}{\mathbf{C}}
\newcommand{\D}{\mathbf{D}}

\newcommand{\G}{\mathbf{G}}
\newcommand{\I}{\mathbf{I}}

\newcommand{\Q}{\mathbf{Q}}

\newcommand{\s}{\mathbf{s}}

\newcommand{\U}{\mathbf{U}}

\newcommand{\X}{\mathbf{X}}
\newcommand{\x}{\mathbf{x}}

\newcommand{\y}{\mathbf{y}}
\newcommand{\GG}{\mathbf{\Gamma}}

\newcommand{\Hc}{\mathbf{H}}

\newcommand{\mLambda}{\mathbf{\Lambda}}

\newcommand{\tr}{\mathrm{Tr}}

\newcommand{\norm}[1]{\mbox{$\left\lVert #1 \right\rVert$}}

\newcommand{\Pwrcon}{\Omega_{+}(P)}

\newcommand{\ddet}[1]{\det\left( #1 \right)}

\newtheorem{thm}{Theorem}

\newcommand{\diag}[1]{\mbox{Diag}\mbox{$\left\{ #1 \right\}$}}

\begin{document}

\title{Maximum Weighted Sum Rate of Multi-Antenna Broadcast Channels}

\author{
\authorblockN{Jia Liu and Y. Thomas Hou}
\authorblockA{ Department of Electrical and Computer Engineering\\
Virginia Polytechnic Institute and State University, Blacksburg, VA 24061
\\Email: \{kevinlau, thou\}@vt.edu}
 }

\maketitle

\begin{abstract}
Recently, researchers showed that dirty paper coding (DPC) is the optimal transmission strategy for
multiple-input multiple-output broadcast channels (MIMO-BC). In this paper, we study how to
determine the maximum weighted sum of DPC rates through solving the maximum weighted sum rate
problem of the dual MIMO multiple access channel (MIMO-MAC) with a sum power constraint. We first
simplify the maximum weighted sum rate problem such that enumerating all possible decoding orders
in the dual MIMO-MAC is unnecessary. We then design an efficient algorithm based on conjugate
gradient projection (CGP) to solve the maximum weighted sum rate problem. Our proposed CGP method
utilizes the powerful concept of Hessian conjugacy. We also develop a rigorous algorithm to solve
the projection problem. We show that CGP enjoys provable convergence, nice scalability, and great
efficiency for large MIMO-BC systems.
\end{abstract}

\section{Introduction}
The capacity region of multiple-input multiple-output broadcast channels (MIMO-BC) has received
great attention in recent years. MIMO-BC belong to the class of nondegraded broadcast channels, for
which the capacity region is notoriously hard to analyze \cite{Cover91:Info_Thry}. Very recently,
researchers have made significant progress in this area. Most notably, Weigarten {\em et. al.}
finally proved the long-open conjecture in \cite{Weingarten06:MIMO_BC} that the ``dirty paper
coding'' (DPC) strategy is the capacity achieving transmission strategy for MIMO-BC. Moreover, by
the remarkable channel {\em duality} between MIMO-BC and its dual MIMO multiple access channel
(MIMO-MAC) \cite{Vishwanath03:duality, Viswanath03:duality, Yu06:duality}, the nonconvex MIMO-BC
capacity region (with respect to the input covariance matrices) can be transformed to the convex
dual MIMO-MAC capacity region with a sum power constraint.

In this paper, we study how to determine the {\em maximum weighted sum of DPC rates} (MWSR) of
MIMO-BC through solving the maximum weighted sum rate problem of the dual MIMO-MAC. Important
applications of the MWSR problem of MIMO-BC include but are not limited to applying Lagrangian dual
decomposition for the cross-layer optimization for MIMO-based mesh networks
\cite{Liu07_CL_MIMO_BC_TR}. The MWSR problem of MIMO-BC is the {\em general} case of the maximum
sum rate problem (MSR) of MIMO-BC, which has been solved by using various algorithms in the
literature. Such algorithms include the minimax method (MM) by Lan and Yu \cite{Lan_Yu_GLOBECOM04},
the steepest descent (SD) method by Viswanathan {\em et al.} \cite{Viswanathan03:MIMO_BC_SD}, the
dual decomposition (DD) method by Yu \cite{Yu_CISS03}, two iterative water-filling methods (IWFs)
by Jindal {\em et al.}\cite{Jindal04:MIMO_BC_IWF}, and the conjugate gradient projection method
recently proposed by us \cite{Liu07:MIMO_BC_CGP}. Among these algorithms, IWFs and CGP appear to be
the simplest. However, all of these existing algorithms have limitations in that they cannot be
readily extended to the MWSR problem of MIMO-BC. As we show later, the objective function of the
MWSR problem has a very different and much more complex objective function. The aforementioned
algorithms can only handle the objective function of MSR, which is just a special case of MWSR (by
setting all weights to one). These limitations of the existing algorithms motivate us to design an
efficient and scalable algorithm with a modest storage requirement to solve the MWSR problem of
large MIMO-BC systems.

In this paper, we significantly extend our CGP method in \cite{Liu07:MIMO_BC_CGP} to handle the
MWSR problem of MIMO-BC. Our CGP method is inspired by \cite{Ye03:MIMO_SH_AdHoc}, where a gradient
projection method was used to heuristically solve the MSR problem of MIMO interference channels.
However, unlike \cite{Ye03:MIMO_SH_AdHoc}, we use the {\em conjugate} gradient directions instead
of gradient directions to eliminate the ``zigzagging'' phenomenon. Also different from
\cite{Ye03:MIMO_SH_AdHoc}, we develop a rigorous algorithm to exactly solve the projection problem.
Our main contributions in this paper are three-fold:
\begin{enumerate}
\item To the best of our knowledge, our paper is the first work that considers the MWSR problem of
MIMO-BC. Studying the MWSR problem is more useful and more important because the MWSR problem is
the general case of MSR, and it has much wider application in systems and networks that employ
MIMO-BC.

\item We simplify the MWSR problem of the dual MIMO-MAC such that enumerating
all different decoding orders in the dual MIMO-MAC is {\em unnecessary}, thus paving the way to
design an algorithm to efficiently solve the MWSR problem of MIMO-BC.

\item We extend the CGP method in \cite{Liu07:MIMO_BC_CGP} for the MWSR problem of MIMO-BC.
This extended CGP method still enjoys provable convergence as well as nice scalability, and has the
desirable linear complexity. Also, the extended CGP method is insensitive to the increase of the
number of users and has a modest memory requirement.
\end{enumerate}

The remainder of this paper is organized as follows. In Section~\ref{sec:model}, we discuss the
network model and the problem formulation. Section~\ref{sec:framework} introduces the key
components in of CGP, including the computation of conjugate gradients and performing projection.
We analyze the complexity of CGP in Section~\ref{sec:complexity}. Numerical results of CGP's
convergence behavior and performance comparison with other existing algorithms are presented in
Section~\ref{sec:results}. Section~\ref{sec:conclusions} concludes this paper.

\section{System Model and Problem Formulation} \label{sec:model}
We begin with introducing notations. We use boldface to denote matrices and vectors. For a
complex-valued matrix $\A$, $\A^{*}$ and $\A^{\dag}$ denote the conjugate and the conjugate
transpose of $\A$, respectively. $\tr\{\A\}$ denotes the trace of $\A$. We let $\I$ denote the
identity matrix with dimension determined from context. $\A \succeq 0$ represents that $\A$ is
Hermitian and positive semidefinite (PSD). $\diag{\A_{1} \ldots \A_{n}}$ denotes the block diagonal
matrix with matrices $\A_{1},\ldots,\A_{n}$ on its main diagonal.

Suppose that a MIMO Gaussian broadcast channel has $K$ users, each of which is equipped with
$n_{r}$ antennas, and the transmitter has $n_{t}$ antennas. The channel matrix for user $i$ is
denoted as $\Hc_{i} \in \mathbb{C}^{n_{r} \times n_{t}}$. In \cite{Weingarten06:MIMO_BC}, it has
been shown that the capacity region of MIMO-BC is equal to the dirty-paper coding region (DPC). In
DPC rate region, suppose that users $1,\ldots,K$ are encoded subsequently, then the rate of user
$i$ can be computed as \cite{Vishwanath03:duality}
\begin{equation} \label{eqn_dpc_rate}
R_{i}^{\mathrm{DPC}}(\GG) = \log \frac{\ddet{\I + \Hc_{i} \left(\sum_{j=i}^{K} \GG_{j} \right)
\Hc_{i}^{\dag} }}{\ddet{\I + \Hc_{i} \left( \sum_{j=i+1}^{K} \GG_{j} \right) \Hc_{i}^{\dag}} },
\end{equation}
where $\GG_{i} \in \mathbb{C}^{n_{t}\times n_{t}}$, $i=1,\ldots,K$, are the {\em downlink} input
covariance matrices, $\GG \triangleq \{\GG_{1},\ldots\GG_{K}\}$ denotes the collection of all the
downlink covariance matrices. As a result, the MWSR problem can then be written as follows:
\begin{equation} \label{eqn_bc_mws}
\begin{array}{rl}
\mbox{Maximize} & \sum_{i=1}^{K} u_{i} R_{i}^{\mathrm{DPC}} (\GG)
\\
\mbox{subject to} & \GG_{i} \succeq 0, \quad i=1,\ldots,K \\
& \sum_{i=1}^{K} \tr (\GG_{i}) \leq P,
\end{array}
\end{equation}
where $u_{i}$ is the weight of user $i$, $P$ represents the maximum transmit power at the
transmitter. It is evident that (\ref{eqn_bc_mws}) is a nonconvex optimization problem since the
DPC rate equation in (\ref{eqn_dpc_rate}) is neither a concave nor a convex function in the input
covariance matrices $\GG_{1},\ldots,\GG_{K}$. However, the authors in \cite{Vishwanath03:duality}
showed that due to the duality between MIMO-BC and MIMO-MAC, the rates achievable in MIMO-BC are
also achievable in MIMO-MAC. That is, given a feasible $\GG$, there exists a set of feasible {\em
uplink} input covariance matrices for the dual MIMO-MAC, denoted by $\Q$, such that
$R_{i}^{\mathrm{MAC}}(\Q) = R_{i}^{\mathrm{DPC}}(\GG)$. Thus, (\ref{eqn_bc_mws}) is equivalent to
the following maximum weighted sum rate problem of the dual MIMO-MAC with a sum power constraint:
\begin{equation} \label{eqn_mac_mws}
\begin{array}{rl}
\hspace{-.1in} \mbox{Maximize} & \sum_{i=1}^{K} u_{i} R_{i}^{\mathrm{MAC}}(\Q) \\
\hspace{-.1in} \mbox{subject to} & \Q_{i} \succeq 0, \quad i=1,\ldots,K\\
\hspace{-.1in} & R_{i}^{\mathrm{MAC}}(\Q) \in \mathcal{C}_{\mathrm{MAC}}(P,\Hc^{\dag}), \,\, i=1,\ldots,K\\
\hspace{-.1in} & \sum_{i=1}^{K} \tr (\Q_{i}) \leq P, \\
\end{array}
\end{equation}
where $\Q_{i} \in \mathbb{C}^{n_{r}\times n_{r}}$, $i=1,\ldots,K$, are the uplink input covariance
matrices, $\Q \triangleq \{\Q_{1},\ldots\Q_{K}\}$ represents the collection of all the uplink
covariance matrices, $\mathcal{C}_{\mathrm{MAC}}(P,\Hc^{\dag})$ represents the capacity region of
the dual MIMO-MAC. It is known that the capacity region of a MIMO-MAC can be achieved by the
successive decoding \cite{Cover91:Info_Thry}. However, in order to determine the capacity region of
a MIMO-MAC, all possible successive decoding orders need to be enumerated, which is very
cumbersome. In the following theorem, however, we show that the enumeration of all successive
decoding orders is indeed {\em unnecessary} when solving the MWSR problem of the dual MIMO-MAC.
This result significantly reduces the complexity and paves the way to efficiently solve the MWSR
problem by using CGP method.

\begin{thm} \label{thm_weighted_sum}
The MWSR problem in (\ref{eqn_mac_mws}) can be solved by the following equivalent optimization
problem:
\begin{equation} \label{eqn_mac_equiv}
\begin{array}{rl}
\!\!\!\!\!\! \mbox{Maximize} & \!\!\! \sum_{i=1}^{K} (u_{\pi(i)} - u_{\pi(i-1)}) \times \\
\!\!\!\!\!\! & \!\!\! \log \ddet{\I + \sum_{j=i}^{K} \Hc_{\pi(j)}^{\dag} \Q_{\pi(j)} \Hc_{\pi(j)} } \\
\!\!\!\!\!\!  \mbox{subject to} & \!\!\! \sum_{i=1}^{K} \tr(\Q_{i}) \leq P_{\max} \\
\!\!\!\!\!\! & \!\!\! \Q_{i} \succeq 0, \,\, i=1,\ldots,K,
\end{array}
\end{equation}
where $u_{\pi(0)} \triangleq 0$, $\pi$ is a permutation of the set $\{1,\ldots,K\}$ such that
$u_{\pi(1)} \leq \ldots \leq u_{\pi(K)}$. $\pi(i), i=1,\ldots,K$, represents the $i^{th}$ position
in permutation $\pi$.
\end{thm}

\begin{proof}
Let $\Phi(\mathcal{S}) = \log \det (\I + \sum_{i\in \mathcal{S}} \Hc_{\pi(i)}^{\dag} \Q_{\pi(i)}
\Hc_{\pi(i)})$, where $\mathcal{S}$ is a non-empty subset of $\{1,\ldots,K\}$. From Theorem 14.3.5
in \cite{Cover91:Info_Thry}, we know that the maximum weighted sum rate problem can be written as
\begin{equation*}
\begin{array}{rl}
\mbox{Maximize} & \sum_{i=1}^{K} u_{\pi(i)} R_{\pi(i)}^{\mathrm{MAC}} \\
\mbox{subject to} & \sum_{i \in \mathcal{S}} R_{\pi(i)}^{\mathrm{MAC}} \leq \Phi(\mathcal{S}), \,\,
\forall \mathcal{S} \subseteq \{1,\ldots,K\}.
\end{array}
\end{equation*}
Thus, it is not difficult to see that, when $\mathcal{S} = \{\pi(i)\}$, $R_{\pi(i)}^{\mathrm{MAC}}
\leq \Phi(\{\pi(i)\}) = \log\ddet{\I + \Hc_{\pi(i)}^{\dag} \Q_{\pi(i)} \Hc_{\pi(i)} }$. Since
$u_{\pi(1)} \leq \ldots \leq u_{\pi(K)}$, from Karush-Kuhn-Tucker (KKT) condition, we must have
that the constraint $R_{\pi(K)}^{\mathrm{MAC}} \leq \Phi(\{ \pi(K) \})$ must be tight at
optimality. That is,
\begin{equation} \label{eqn_R_K}
R_{\pi(K)}^{\mathrm{MAC}} = \log \ddet {\I + \Hc_{\pi(K)}^{\dag} \Q_{\pi(K)} \Hc_{\pi(K)} }.
\end{equation}
Likewise, when $\mathcal{S} = \{\pi(K-1),\pi(K)\}$, we have
\begin{equation*}
\begin{array}{l}
\hspace{-.2in} R_{\pi(K-1)}^{\mathrm{MAC}} + R_{\pi(K)}^{\mathrm{MAC}} \leq \log \det \left( \I
+ \Hc_{\pi(K)}^{\dag} \Q_{\pi(K)} \Hc_{\pi(K)} \right. \\
\hspace{+1in} \left.+ \Hc_{\pi(K-1)}^{\dag} \Q_{\pi(K-1)} \Hc_{\pi(K-1)} \right).
\end{array}
\end{equation*}
So, from (\ref{eqn_R_K}), we have
\begin{equation} \label{eqn_thm3_2}
\begin{array}{l}
\hspace{-.0in} R_{\pi(K-1)}^{\mathrm{MAC}} \leq \log \det\left( \I + \Hc_{\pi(K)}^{\dag} \Q_{\pi(K)} \Hc_{\pi(K)} \right. \\
\hspace{+0.8in} \left. + \Hc_{\pi(K-1)}^{\dag} \Q_{\pi(K-1)} \Hc_{\pi(K-1)} \right) - \\
\hspace{+0.8in} \log \ddet{\I + \Hc_{\pi(K)}^{\dag} \Q_{\pi(K)} \Hc_{\pi(K)} }.
\end{array}
\end{equation}
Since $u_{\pi(K-1)}$ is the second largest weight, again from KKT condition, we must have that
(\ref{eqn_thm3_2}) must be tight at optimality. This process continues for all $K$ users.
Subsequently, we have that
\begin{equation} \label{eqn_thm3_3}
\begin{array}{l}
\hspace{-.1in} R_{\pi(i)}^{\mathrm{MAC}} = \log \ddet{\I + \sum_{j=i}^{K} \Hc_{\pi(j)}^{\dag}
\Q_{\pi(j)} \Hc_{\pi(j)} } \\
\hspace{.4in} - \log \ddet{\I + \sum_{j=i+1}^{K} \Hc_{\pi(j)}^{\dag} \Q_{\pi(j)} \Hc_{\pi(j)} },
\end{array}
\end{equation}
for $i=1,\ldots,K-1$. Summing up all $u_{\pi(i)} R_{\pi(i)}^{\mathrm{MAC}}$ and after rearranging
the terms, it is readily verifiable that
\begin{equation} \label{eqn_thm3_4}
\begin{array}{l}
\hspace{-.1in} \sum_{i=1}^{K} u_{\pi(i)} R_{\pi(i)}^{\mathrm{MAC}} = \sum_{i=1}^{K} (u_{\pi(i)} - u_{\pi(i-1)}) \times \nonumber \\
\hspace{1in} \log \ddet{\I + \sum_{j=i}^{K} \Hc_{\pi(j)}^{\dag} \Q_{\pi(j)} \Hc_{\pi(j)} }.
\end{array}
\end{equation}
It then follows that the MWSR problem of the dual MIMO-MAC is equivalent to maximizing
(\ref{eqn_thm3_4}) with the sum power constraint, i.e., the optimization problem in
(\ref{eqn_mac_equiv}).
\end{proof}

An important observation from (\ref{eqn_mac_equiv}) is that, since $\log\ddet{\cdot}$ is a concave
function for positive semidefinite matrices \cite{Cover91:Info_Thry}, (\ref{eqn_mac_equiv}) is a
convex optimization problem with respect to the uplink input covariance matrices
$\Q_{\pi(1)},\ldots,\Q_{\pi(K)}$. However, although the standard interior point convex optimization
method can be used to solve (\ref{eqn_mac_equiv}), it is considerably more complex than a method
that exploits the special structure of (\ref{eqn_mac_equiv}).

\section{Conjugate Gradient Projection Method} \label{sec:framework}
In this paper, we modified the conjugate gradient projection method (CGP) in
\cite{Liu07:MIMO_BC_CGP} to solve (\ref{eqn_mac_equiv}). CGP utilizes the important and powerful
concept of Hessian conjugacy to deflect the gradient direction appropriately so as to achieve the
superlinear convergence rate \cite{Bazaraa_Sherali_Shetty_93:NLP}. The framework of CGP for solving
(\ref{eqn_mac_equiv}) is shown in Algorithm~\ref{alg_mgp}.
\begin{algorithm}
{\footnotesize \caption{Gradient Projection Method} \label{alg_mgp}
\begin{algorithmic}
\STATE {\bf Initialization:} \\
\STATE \quad Choose the initial conditions $\Q^{(0)} = [ \Q_{1}^{(0)}, \Q_{2}^{(0)}, \ldots, \Q_{K}^{(0)}]^{T}$. Let \\
\STATE \quad $k=0$. \\
\STATE {\bf Main Loop:} \\
\STATE \quad 1. Calculate the conjugate gradients $\G_{i}^{(k)}$, $i= 1,2, \ldots, K$. \\
\STATE \quad 2. Choose an appropriate step size $s_{k}$. Let $\Q_{i}^{'(k)} = \Q_{i}^{(k)} + s_{k}
\G_{i}^{(k)}$, \\
\STATE \quad \quad for $i=1,2,\ldots,K$. \\
\STATE \quad 3. Let $\bar{\Q}^{(k)}$ be the projection of $\Q^{'(k)}$ onto $\Pwrcon$, where
$\Pwrcon \triangleq $\\
\STATE \quad \quad $\{ \Q_{i}, \,\, i=1,\ldots,K | \Q_{i} \succeq 0, \sum_{i=1}^{K} \tr\{ \Q_{i} \} \leq P \}$. \\
\STATE \quad 4. Choose appropriate step size $\alpha_{k}$. Let $\Q_{l}^{(k+1)} = \Q_{l}^{(k)} +
\alpha_{k}(\bar{\Q}_{i}^{(k)} - $ \\
\STATE \quad \quad $\Q_{i}^{(k)})$, $i=1,2,\ldots,K$. \\
\STATE \quad 5. $k=k+1$. If the maximum absolute value of the elements in $\Q_{i}^{(k)} - $\\
\STATE \quad \quad $\Q_{i}^{(k-1)}< \epsilon$, for $i=1,2,\ldots,L$, then stop; else go to step 1.
\end{algorithmic}}
\end{algorithm}

Due to the complexity of the objective function in (\ref{eqn_mac_equiv}), we adopt the inexact line
search method called ``Armijo's Rule'' to avoid excessive objective function evaluations, while
still enjoying provable convergence \cite{Bazaraa_Sherali_Shetty_93:NLP}. The basic idea of
Armijo's Rule is that at each step of the line search, we sacrifice accuracy for efficiency as long
as we have sufficient improvement. According to Armijo's Rule, in the $k^{th}$ iteration, we choose
$\sigma_{k}=1$ and $\alpha_{k} = \beta^{m_{k}}$ (the same as in \cite{Ye03:MIMO_SH_AdHoc}), where
$m_{k}$ is the first non-negative integer $m$ that satisfies
\begin{eqnarray} \label{eqn_Armijo}
\!\!\!\!\!\! && F(\Q^{(k+1)}) - F(\Q^{(k)})\geq \sigma \beta^{m} \langle
\G^{(k)}, \bar{\Q}^{(k)} - \Q^{(k)} \rangle \nonumber\\
\!\!\!\!\!\! && = \sigma \beta^{m} \sum_{i=1}^{K} \tr \left[ \G_{i}^{\dag(k)}
\left(\bar{\Q}_{i}^{(k)} - \Q_{i}^{(k)}\right) \right],
\end{eqnarray}
where $0 < \beta < 1$ and $0 < \sigma < 1$ are fixed scalars.

Next, we will consider two major components in the CGP framework: 1) how to compute the conjugate
gradient direction $\G_{i}$, and 2) how to project $\Q^{'(k)}$ onto the set $\Pwrcon \triangleq \{
\Q_{i}, \,\, i=1,\ldots,K | \Q_{i} \succeq 0, \sum_{i=1}^{K} \tr\{ \Q_{i} \} \leq P \}$.

\subsection{Computing the Conjugate Gradients} The gradient $\bar{\G}_{\pi(j)} \triangleq \nabla_{\Q_{\pi(j)}}
F(\Q)$ depends on the partial derivative of $F(\Q)$ with respect to $\Q_{\pi(j)}$. By using the
formula $\frac{\partial \ln \ddet{\A+\B\X\C}}{\partial \X} = \left[ \C(\A+\B\X\C)^{-1}\B
\right]^{T}$ \cite{Ye03:MIMO_SH_AdHoc, Magnus_Neudecker99:Mtrx_Diff_Calc}, we can compute the
partial derivative of the $i^{th}$ term in the summation of $F(\Q)$ with respect to $\Q_{\pi(j)}$,
$j \geq i$, as follows:
\begin{eqnarray*} \label{eqn_partial}
&& \hspace{-.25in}\frac{\partial} {\partial \Q_{\pi(j)}} \Bigg( (u_{\pi(i)} - u_{\pi(i-1)}) \times
\nonumber \\
&& \hspace{.3in} \left. \log \ddet { \I + \sum_{k=i}^{K} \Hc_{\pi(k)}^{\dag} \Q_{\pi(k)} \Hc_{\pi(k)} } \right) \nonumber \\
&& \hspace{-.25in} = \left(u_{\pi(i)} - u_{\pi(i-1)}\right) \times \nonumber\\ && \hspace{-.3in}
\left[ \Hc_{\pi(j)} \left( \I + \sum_{k=i}^{K} \Hc_{\pi(k)}^{\dag} \Q_{\pi(k)} \Hc_{\pi(k)}
\right)^{-1} \Hc_{\pi(j)}^{\dag} \right]^{T} \hspace{-.1in}.
\end{eqnarray*}
To compute the gradient of $F(\Q)$ with respect to $\Q_{\pi(j)}$, we notice that only the first $j$
terms in $F(\Q)$ involve $\Q_{\pi(j)}$. From the definition $\nabla_{z} f(z) = 2(\partial
f(z)/\partial z)^{*}$ \cite{Haykin96:Adpt_Fltr}, we have
\begin{eqnarray} \label{eqn_gradient}
&& \hspace{-.25in} \bar{\G}_{\pi(j)} = 2\Hc_{\pi(j)} \Bigg[ \sum_{i=1}^{j}\left(
u_{\pi(i)} - u_{\pi(i-1)} \right) \times \nonumber\\
&& \hspace{-.2in} \left. \left( \I + \sum_{k=i}^{K} \Hc_{\pi(k)}^{\dag} \Q_{\pi(k)} \Hc_{\pi(k)}
\right)^{-1} \right] \Hc_{\pi(j)}^{\dag}.
\end{eqnarray}

It is worth to point out that we can exploit the special structure in (\ref{eqn_gradient}) to
significantly reduce the computation complexity in the implementation of the algorithm. Note that
the most difficult part in computing $\bar{\G}_{\pi(j)}$ is the summation of the terms in the form
of $\Hc_{\pi(k)}^{\dag} \Q_{\pi(k)}\Hc_{\pi(k)}$. Without careful consideration, one may end up
computing such additions $j(2K+1-j)/2$ times for $\bar{\G}_{\pi(j)}$. However, noting that most of
the terms in the summation are still the same when $j$ varies, we can maintain a running sum for
$\I + \sum_{k=i}^{K} \Hc_{\pi(k)}^{\dag} \Q_{\pi(k)} \Hc_{\pi(k)}$, start out from $j=K$, and
reduce $j$ by one sequentially. As a result, only one new term is added to the running sum in each
iteration, which means we only need to do the addition once in each iteration.

The conjugate gradient direction in the $m^{th}$ iteration can be computed as $\G_{\pi(j)}^{(m)} =
\bar{\G}_{\pi(j)}^{(m)} + \rho_{m} \G_{\pi(j)}^{(m-1)}$. We adopt the Fletcher and Reeves' choice
of deflection \cite{Bazaraa_Sherali_Shetty_93:NLP}, which can be computed as
\begin{equation} \label{eqn_conjuate}
\rho_{m} = \frac{\| \bar{\G}_{\pi(j)}^{(m)} \|^{2}}{\| \bar{\G}_{\pi(j)}^{(m-1)} \|^{2}}.
\end{equation}
The purpose of deflecting the gradient using (\ref{eqn_conjuate}) is to find $\G_{\pi(j)}^{(m)}$,
which is the Hessian-conjugate of $\G_{\pi(j)}^{(m-1)}$. By doing so, we can eliminate the
``zigzagging'' phenomenon encountered in the conventional gradient projection method, and achieve
the superlinear convergence rate \cite{Bazaraa_Sherali_Shetty_93:NLP} without actually storing a
large Hessian approximation matrix as in quasi-Newton methods.

\subsection{Projection onto $\Pwrcon$} Noting from (\ref{eqn_gradient}) that $\G_{i}$ is
Hermitian, we have that $\Q_{i}^{'(k)} = \Q_{i}^{(k)}+s_{k} \G_{i}^{(k)}$ is Hermitian as well.
Then, the projection problem becomes how to simultaneously project a set of $K$ Hermitian matrices
onto the set $\Pwrcon$, which contains a constraint on sum power for all users. This is different
to \cite{Ye03:MIMO_SH_AdHoc}, where the projection was performed on individual power constraint. In
order to do this, we construct a block diagonal matrix $\D = \diag{\Q_{1} \ldots \Q_{K}} \in
\mathbb{C}^{(K\cdot n_{r})\times(K\cdot n_{r})}$. It is easy to recognize that $\Q_{i} \in
\Pwrcon$, $i=1,\ldots,K$, only if $\tr(\D) = \sum_{i=1}^{K} \tr \left(\Q_{i}\right) \leq P$ and $\D
\succeq 0$. In this paper, we use Frobenius norm, denoted by $\|\cdot\|_{F}$, as the matrix
distance metric. The distance between two matrices $\A$ and $\B$ is defined as $\| \A - \B \|_{F} =
\left( \tr\left[ (\A-\B)^{\dag} (\A-\B) \right] \right)^{\frac{1}{2}}$. Thus, given a block
diagonal matrix $\D$, we wish to find a matrix $\tilde{\D} \in \Pwrcon$ such that $\tilde{\D}$
minimizes $\| \tilde{\D} - \D \|_{F}$. For more convenient algebraic manipulations, we instead
study the following equivalent optimization problem:
\begin{equation} \label{eqn_proj_primal_equiv}
\begin{array}{rl}
\mbox{Minimize} & \frac{1}{2} \| \tilde{\D} - \D \|_{F}^{2} \\
\mbox{subject to} & \tr (\tilde{\D}) \leq P, \,\, \tilde{\D} \succeq 0. \\
\end{array}
\end{equation}
In (\ref{eqn_proj_primal_equiv}), the objective function is convex in $\tilde{\D}$, the constraint
$\tilde{\D} \succeq 0$ represents the convex cone of positive semidefinite matrices, and the
constraint $\tr (\tilde{\D}) \leq P$ is a linear constraint. Thus, the problem is a convex
minimization problem and we can exactly solve this problem by solving its Lagrangian dual problem.
Associating Hermitian matrix $\X$ to the constraint $\tilde{\D} \succeq 0$ and $\mu$ to the
constraint $\tr (\tilde{\D}) \leq P$, we can write the Lagrangian as
\begin{eqnarray} \label{eqn_prj_lagrangian}
g(\X, \mu) &=& \min_{\tilde{\D}} \left\{ (1/2) \| \tilde{\D} - \D \|_{F}^{2} -
\tr(\X^{\dag} \tilde{\D}) \right.\nonumber\\
&& \quad \quad \,\,+ \left. \mu \left(\tr(\tilde{\D})-P \right) \right\}.
\end{eqnarray}
Since $g(\X, \mu)$ is an unconstrained quadratic minimization problem, we can compute the minimizer
of (\ref{eqn_prj_lagrangian}) by simply setting the derivative of (\ref{eqn_prj_lagrangian}) (with
respect to $\tilde{\D}$) to zero, i.e., $(\tilde{\D} - \D) - \X^{\dag} + \mu \I = 0$. Noting that
$\X^{\dag} = \X$, we have $\tilde{\D} = \D - \mu \I + \X$. Substituting $\tilde{\D}$ back into
(\ref{eqn_prj_lagrangian}), we have
\begin{eqnarray}
&& \!\!\!\!\!\!\!\!\!\!\!\!\!g(\X,\mu) = \frac{1}{2} \norm{ \X - \mu\I }_{F}^{2} - \mu P + \tr \left[ \left(\mu\I - \X \right) \left(\D + \X -\mu\I \right) \right] \nonumber\\
&& \quad \,\, = -\frac{1}{2} \norm{ \D - \mu\I + \X }_{F}^{2} - \mu P + \frac{1}{2} \| \D
\|_{F}^{2}.
\end{eqnarray}
Therefore, the Lagrangian dual problem can be written as
\begin{equation}\label{eqn_prj_dual}
\begin{array}{rl}
\mbox{Maximize} & -\frac{1}{2} \norm{ \D - \mu\I + \X }_{F}^{2} - \mu P + \frac{1}{2} \| \D \|_{F}^{2} \\
\mbox{subject to} & \X \succeq 0, \mu \geq 0.
\end{array}
\end{equation}
After solving (\ref{eqn_prj_dual}), we can have the optimal solution to
(\ref{eqn_proj_primal_equiv}) as:
\begin{equation}
\tilde{\D}^{*} = \D - \mu^{*} \I + \X^{*},
\end{equation}
where $\mu^{*}$ and $\X^{*}$ are the optimal dual solutions to Lagrangian dual problem in
(\ref{eqn_prj_dual}). Although the Lagrangian dual problem in (\ref{eqn_prj_dual}) has a similar
structure as that in the primal problem in (\ref{eqn_proj_primal_equiv}) (having a positive
semidefinitive matrix constraint), we find that the positive semidefinite matrix constraint can
indeed be easily handled. To see this, we first introduce Moreau Decomposition Theorem from convex
analysis.
\begin{thm}\label{thm_moreau}(Moreau Decomposition \cite{Hiriart-Urruty_Lemarechal01:Cnvx_Anl})
Let $\mathcal{K}$ be a closed convex cone. For $\x,\x_{1},\x_{2} \in \mathbb{C}^{p}$, the two
properties below are equivalent:
\begin{enumerate}
\item $\x = \x_{1} + \x_{2}$ with $\x_{1} \in \mathcal{K}$, $\x_{2} \in \mathcal{K}^{o}$ and $\langle \x_{1},\x_{2} \rangle =
0$,
\item $\x_{1} = p_{\mathcal{K}}(\x)$ and $\x_{2} = p_{\mathcal{K}^{o}}(x)$,
\end{enumerate}
where $\mathcal{K}^{o} \triangleq \{ \s \in \mathbb{C}^{p}: \langle \s, \y \rangle \leq 0, \,
\forall \, \y \in \mathcal{K} \}$ is called the polar cone of cone $\mathcal{K}$,
$p_{\mathcal{K}}(\cdot)$ represents the projection onto cone $\mathcal{K}$.
\end{thm}

In fact, the projection onto a cone $\mathcal{K}$ is analogous to the projection onto a subspace.
The only difference is that the orthogonal subspace is replaced by the polar cone.

Now we consider how to project a Hermitian matrix $\A \in \mathbb{C}^{n \times n}$ onto the
positive and negative semidefinite cones. First, we can perform eigenvalue decomposition on $\A$
yielding $\A = \hat{\U} \diag{\lambda_{i}, \,\, i=1,\ldots,n } \hat{\U}^{\dag}$, where $\hat{\U}$
is the unitary matrix formed by the eigenvectors corresponding to the eigenvalues $\lambda_{i}$,
$i=1,\ldots,n$. Then, we have the positive semidefinite and negative semidefinite projections of
$\A$ as follows:
\begin{eqnarray}
\label{eqn_prj_psd} &\A_{+} = \hat{\U} \diag{ \max\{ \lambda_{i},0\}, i=1,2,\ldots,n } \hat{\U}^{\dag},& \\
\label{eqn_prj_nsd} &\A_{-} = \hat{\U} \diag{ \min\{ \lambda_{i},0\}, i=1,2,\ldots,n }
\hat{\U}^{\dag}.&
\end{eqnarray}
The proof of (\ref{eqn_prj_psd}) and (\ref{eqn_prj_nsd}) is a straightforward application of
Theorem~\ref{thm_moreau} by noting that $\A_{+} \succeq 0$, $\A_{-} \preceq 0$, $\langle \A_{+},
\A_{-} \rangle = 0$, $\A_{+} + \A_{-} = \A$, and the positive semidefinite cone and negative
semidefinite cone are polar cones to each other.

We now consider the term $\D - \mu\I + \X$, which is the only term involving $\X$ in the dual
objective function. We can rewrite it as $\D - \mu\I - (-\X)$, where we note that $-\X \preceq 0$.
Finding a negative semidefinite matrix $-\X$ such that $\| \D - \mu\I - (-\X) \|_{F}$ is minimized
is equivalent to finding the projection of $\D - \mu\I$ onto the negative semidefinite cone. From
the previous discussion, we immediately have
\begin{equation} \label{eqn_prj_negX}
-\X = \left(\D - \mu\I \right)_{-}.
\end{equation}
Since $\D - \mu \I = (\D - \mu \I)_{+} + (\D - \mu \I)_{-}$, substituting (\ref{eqn_prj_negX}) back
to the Lagrangian dual objective function, we have
\begin{equation}
\min_{\X} \norm{\D - \mu\I + \X}_{F} = \left(\D - \mu\I \right)_{+}.
\end{equation}
Thus, the matrix variable $\X$ in the Lagrangian dual problem can be removed and the Lagrangian
dual problem can be rewritten as
\begin{equation}\label{eqn_prj_dual_sim}
\begin{array}{rl}
\!\!\!\!\!\!\! \mbox{Maximize} & \!\!\! \psi(\mu) \triangleq -\frac{1}{2} \norm{ \left(\D - \mu\I
\right)_{+} }_{F}^{2} - \mu P + \frac{1}{2} \norm{ \D }_{F}^{2} \\
\!\!\!\!\!\!\! \mbox{subject to} & \!\!\! \mu \geq 0.
\end{array}
\end{equation}
Suppose that after performing eigenvalue decomposition on $\D$, we have $\D = \U \mLambda
\U^{\dag}$, where $\mLambda$ is the diagonal matrix formed by the eigenvalues of $\D$, $\U$ is the
unitary matrix formed by the corresponding eigenvectors. Since $\U$ is unitary, we have $\left(\D -
\mu\I \right)_{+} = \U \left(\mLambda - \mu \I \right)_{+} \U^{\dag}$. It then follows that
\begin{equation}
\norm{ \left(\D - \mu\I \right)_{+} }_{F}^{2} = \norm{\left(\mLambda - \mu \I \right)_{+}}_{F}^{2}.
\end{equation}
We denote the eigenvalues in $\mLambda$ by $\lambda_{i}$, $i=1,2,\ldots,K\cdot n_{r}$. Suppose that
we sort them in non-increasing order such that $\mLambda = \diag{\lambda_{1} \,\, \lambda_{2}
\ldots \,\, \lambda_{K\cdot n_{r}}}$, where $\lambda_{1} \geq \ldots \geq \lambda_{K\cdot n_{r}}$.
It then follows that
\begin{equation} \label{eqn_norm}
\norm{\left(\mLambda - \mu \I \right)_{+}}_{F}^{2} = \sum_{j=1}^{K \cdot n_{r}} \left( \max
\left\{0, \lambda_{j} - \mu \right\} \right)^{2}.
\end{equation}
From (\ref{eqn_norm}), we can rewrite $\psi(\mu)$ as
\begin{equation} \label{eqn_psi}
\psi(\mu) = - \frac{1}{2} \sum_{j=1}^{K \cdot n_{r}} \left( \max \left\{0, \lambda_{j} - \mu
\right\} \right)^{2} - \mu P + \frac{1}{2} \norm{\D}_{F}^{2}.
\end{equation}
It is evident from (\ref{eqn_psi}) that $\psi(\mu)$ is continuous and (piece-wise) concave in
$\mu$. Generally, piece-wise concave maximization problems can be solved by using the subgradient
method. However, due to the heuristic nature of its step size selection strategy, subgradient
algorithm usually does not perform well. In fact, by exploiting the special structure,
(\ref{eqn_prj_dual_sim}) can be efficiently solved. We can search the optimal value of $\mu$ as
follows. Let $\hat{I}$ index the pieces of $\psi(\mu)$, $\hat{I}=0,1,\ldots,K\cdot n_{r}$.
Initially we set $\hat{I}=0$ and increase $\hat{I}$ subsequently. Also, we introduce $\lambda_{0} =
\infty$ and $\lambda_{K\cdot n_{r}+1} = -\infty$. We let the endpoint objective value
$\psi_{\hat{I}}\left(\lambda_{0} \right) = 0$, $\phi^{*} = \psi_{\hat{I}} \left( \lambda_{0}
\right)$, and $\mu^{*} = \lambda_{0}$. If $\hat{I} > K \cdot n_{r}$, the search stops. For a
particular index $\hat{I}$, by setting
\begin{equation}
\frac{\partial}{\partial \mu} \psi_{\hat{I}}(\nu) \triangleq \frac{\partial}{\partial \mu} \left(
-\frac{1}{2} \sum_{i=1}^{\hat{I}}\left(\lambda_{i}-\mu \right)^{2} - \mu P \right) = 0,
\end{equation}
we have
\begin{equation}
\mu_{\hat{I}}^{*} = \frac{\sum_{i=1}^{\hat{I}} \lambda_{i} - P }{\hat{I}}.
\end{equation}
Now we consider the following two cases:
\begin{enumerate}
\item If $\mu_{\hat{I}}^{*} \in \left[\lambda_{\hat{I}+1}, \lambda_{\hat{I}}\right] \cap
\mathbb{R}_{+}$, where $\mathbb{R}_{+}$ denotes the set of non-negative real numbers, then we have
found the optimal solution for $\mu$ because $\psi(\mu)$ is concave in $\mu$. Thus, the point
having zero-value first derivative, if exists, must be the unique global maximum solution. Hence,
we can let $\mu^{*}=\mu_{\hat{I}}^{*}$ and the search is done.

\item If $\mu_{\hat{I}}^{*} \notin \left[\lambda_{\hat{I}+1},\lambda_{\hat{I}}\right] \cap \mathbb{R}_{+}$,
we must have that the local maximum in the interval $\left[\lambda_{\hat{I}+1},
\lambda_{\hat{I}}\right] \cap \mathbb{R}_{+}$ is achieved at one of the two endpoints. Note that
the objective value $\psi_{\hat{I}}\left( \lambda_{\hat{I}} \right)$ has been computed in the
previous iteration because from the continuity of the objective function, we have
$\psi_{\hat{I}}\left( \lambda_{\hat{I}} \right) = \psi_{\hat{I}-1}\left( \lambda_{\hat{I}}
\right)$. Thus, we only need to compute the other endpoint objective value
$\psi_{\hat{I}}\left(\lambda_{\hat{I}+1}\right)$. If
$\psi_{\hat{I}}\left(\lambda_{\hat{I}+1}\right) < \psi_{\hat{I}}\left(\lambda_{\hat{I}}\right) =
\phi^{*}$, then we know $\mu^{*}$ is the optimal solution; else let $\mu^{*} =
\lambda_{\hat{I}+1}$, $\phi^{*} = \psi_{\hat{I}}\left( \lambda_{\hat{I}+1} \right)$, $\hat{I} =
\hat{I} + 1$ and continue.
\end{enumerate}
Since there are $K \cdot n_{r}+1$ intervals in total, the search process takes at most $K \cdot
n_{r}+1$ steps to find the optimal solution $\mu^{*}$. Hence, this search is of polynomial-time
complexity $O(n_{r}K)$.

After finding $\mu^{*}$, we can compute $\tilde{\D}^{*}$ as
\begin{equation}
\tilde{\D}^{*} = \left( \D - \mu^{*} \I \right)_{+} = \U \left( \mLambda - \mu^{*} \I \right)_{+}
\U^{\dag}.
\end{equation}
That is, the projection $\tilde{\D}$ can be computed by adjusting the eigenvalues of $\D$ using
$\mu^{*}$ and keeping the eigenvectors unchanged. The projection of $\D$ onto $\Pwrcon$ is
summarized in Algorithm~\ref{alg_rlt_prj}.
\begin{algorithm}
\caption{Projection onto $\Pwrcon$} \label{alg_rlt_prj}
\begin{algorithmic}
{\footnotesize
\STATE {\bf Initiation:} \\
\STATE \quad 1. Construct a block diagonal matrix $\D$. Perform eigenvalue decompo- \\
\STATE \quad \quad sition $\D = \U \mathbf{\Lambda} \U^{\dag}$, sort the eigenvalues in non-increasing order. \\
\STATE \quad 2. Introduce $\lambda_{0} = \infty$ and $\lambda_{K \cdot n_{t}+1} = -\infty$. Let
$\hat{I}=0$. Let the \\
\STATE \quad \quad endpoint objective value $\psi_{\hat{I}}\left(\lambda_{0}\right) = 0$, $\phi^{*}
= \psi_{\hat{I}}\left(\lambda_{0}\right)$, and $\mu^{*} = \lambda_{0}$.
\STATE {\bf Main Loop:} \\
\STATE \quad 1. If $\hat{I} > K \cdot n_{r}$, go to the final step; else let
$\mu_{\hat{I}}^{*} = (\sum_{j=1}^{\hat{I}} \lambda_{j} - P )/\hat{I}$. \\
\STATE \quad 2. If {$\mu_{\hat{I}}^{*} \in [\lambda_{\hat{I}+1}, \lambda_{\hat{I}} ] \cap
\mathbb{R}_{+}$}, then let $\mu^{*} = \mu_{\hat{I}}^{*}$ and go to the final step. \\
\STATE \quad 3. Compute $\psi_{\hat{I}}(\lambda_{\hat{I}+1})$. If $\psi_{\hat{I}}(\lambda_{\hat{I}+1}) < \phi^{*}$, then go to the final step; \\
\STATE \quad \quad else let $\mu^{*} = \lambda_{\hat{I}+1}$, $\phi^{*} = \psi_{\hat{I}}(\lambda_{\hat{I}+1})$, $\hat{I} = \hat{I} + 1$ and continue. \\
\STATE {\bf Final Step:} Compute $\tilde{\D}$ as $\tilde{\D} = \U \left( \mLambda - \mu^{*} \I
\right)_{+} \U^{\dag}$.
 }
\end{algorithmic}
\end{algorithm}

\section{Complexity Analysis} \label{sec:complexity}
In this section, we analyze the complexity of our proposed CGP algorithm. Similar to IWFs
\cite{Jindal04:MIMO_BC_IWF}, SD \cite{Viswanathan03:MIMO_BC_SD}, and DD \cite{Yu_CISS03}, CGP has
the desirable ``linear complexity property''. We list the complexity per iteration for each
component of CGP in Table~\ref{tab_complexity}.
\begin{table}[h!]
\centering {\scriptsize \caption{\label{tab_complexity}Per Iteration Complexity in the Components
of CGP}
\begin{tabular}{|c|c|}
\hline  & CGP \\
\hline Gradient & $K$ \\
\hline Line Search & $O(mK)$ \\
\hline Projection & $O(n_{r}K)$ \\
\hline \hline Overall & $O((m+1+n_{r})K)$ \\
\hline
\end{tabular}}
\end{table}
In CGP, it can be seen that the most time-consuming part (increasing with respect to $K$) is the
addition of the terms in the form of $\Hc_{i}^{\dag} \Q_{i} \Hc_{i}$ when computing gradients.
Since the term $(\I + \sum_{k=i}^{K}\Hc_{i}^{\dag} \Q_{i} \Hc_{i})$ can be computed by the running
sum, we only need to compute this sum once in each iteration. Thus, the number of such additions
per iteration for CGP is $K$. It is also obvious that the projection in each iteration of CGP has
the complexity of $O(n_{r}K)$. The complexity of the Armijo's rule inexact line search has the
complexity of $O(mK)$ (in terms of the additions of $\Hc_{i}^{\dag} \Q_{i} \Hc_{i}$ terms), where
$m$ is the number of trials in Armijo's Rule. Therefore, the overall complexity per iteration for
CGP is $O((m+1+n_{r})K)$. According to our computational experience, the value of $m$ usually lies
in between two and four. This shows that CGP has the linear complexity in $K$.

Also, as evidenced in the next section, the numbers of iterations required for convergence in CGP
is very insensitive to the increase of the number of users. Moreover, CGP has a modest memory
requirement: It only requires the solution information from the previous step, as opposed to the
IWFs, which requires previous $K-1$ steps.

\section{Numerical Results} \label{sec:results}
\begin{figure}[ht!]
\centering
\includegraphics[width=2.8in]{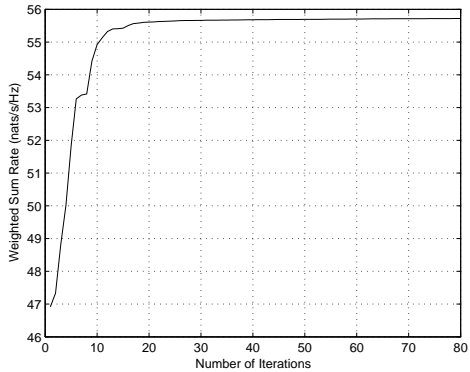}
\caption{Convergence behavior of a 10-user MIMO-BC with $n_{t}=n_{r}=4$.} \label{fig_example3}
\end{figure}

We first use an example of a MIMO-BC system consisting of 10 users with $n_{t}=n_{r}=4$ to show the
convergence behavior of our proposed algorithm. The weights of the 10 users are $[1, 1.5, 0.8, 0.9,
1.4, 1.2, 0.7, 1.1, 1.03, 1.3]$, respectively. The convergence process is plotted in
Fig.~\ref{fig_example3}. It can be seen that CGP takes approximately 30 iterations to reach near
the optimal.

\begin{figure}[ht!]
\centering
\includegraphics[width=2.8in]{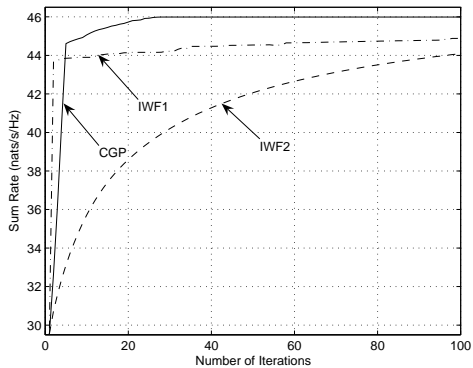}
\caption{Comparison in a 100-user MIMO-BC channel with $n_{t}=n_{r}=4$.} \label{fig_example6}
\end{figure}

To compare the efficiency of CGP with that of IWFs, we give an example of an equal-weight large
MIMO-BC system consisting of 100 users with $n_{t}=n_{r}=4$ in here. The convergence processes are
plotted in Fig.~\ref{fig_example6}. It is observed from Fig.~\ref{fig_example6} that CGP takes only
29 iterations to converge and it outperforms both IWFs. IWF1's convergence speed significantly
drops after the quick improvement in the early stage. It is also seen in this example that IWF2's
performance is inferior to IWF1, and this observation is in accordance with the results in
\cite{Jindal04:MIMO_BC_IWF}. Both IWF1 and IWF2 fail to converge within 100 iterations. The
scalability problem of both IWFs is not surprising because in both IWFs, the most recently updated
covariance matrices only account for a fraction of $1/K$ in the effective channels' computation,
which means it does not effectively make use of the most recent solution. In all of our numerical
examples with different number of users, CGP always converges within 30 iterations.

\section{Conclusion} \label{sec:conclusions}
In this paper, we studied the maximum weighted sum rate (MWSR) problem of MIMO-BC. Specifically, we
derived the MWSR problem of the dual MIMO-MAC with a sum power constraint and developed an
efficient algorithm based on conjugate gradient projection (CGP) to solve the MWSR problem. Also,
we theoretically and numerically analyzed its complexity and convergence behavior. Our
contributions in this paper are three-fold: First, this paper is the first work that considers the
MWSR problem of MIMO-BC; Second, we simplified the MWSR problem in the dual MIMO-MAC and showed
that enumerating all different decoding orders is unnecessary; Third, we developed an efficient and
well-scalable algorithm based on conjugate gradient projection (CGP). The attractive features of
CGP and encouraging results in this paper showed that CGP is an excellent method for solving the
MWSR problem of large MIMO-BC systems.


\end{document}